\documentclass[journal=jacsat,manuscript=article]{achemso}

\usepackage[version=3]{mhchem} 
\usepackage{hyperref}
\usepackage{float}
\usepackage{amsmath}
\usepackage{xcolor}
\usepackage{textcomp}

\author{Anita Pahi}
\email{ap19rs045@iiserkol.ac.in}
\author{Kirty Ranjan Sahoo}
\author{Biswajit Das}
\email{bd18ip005@iiserkol.ac.in}
\affiliation{Indian Institute of Science Education and Research, Kolkata 741246, India}
\author{Shuvojit Paul}
\affiliation{Kandi Raj College, Kandi, Murshidabad 742161, India}

\author{Ayan Banerjee}
\affiliation{Indian Institute of Science Education and Research, Kolkata 741246, India}
\email{ayan@iiserkol.ac.in}

\title[\textsf{achemso} demo]
 {Simultaneous active and diffusive behaviour of asymmetric microclusters in a photophoretic trap}

\abbreviations{}

\begin{document}


\begin{abstract}

Active and diffusive motion in Brownian particles are regularly observed in fluidic environments, albeit at different time scales. Here, we experimentally study the dynamics of highly asymmetric microclusters trapped in air employing photophoretic forces generated from a loosely focused laser beam, where the trapped particles display active and diffusive dynamics simultaneously in orthogonal spatial directions. Thus, particle motion in the longitudinal direction ($z$) is enslaved to irregular kicks that naturally arise from an interplay of gravitational and photophoretic forces. This leads to a bimodal nature of the probability distribution function with a near-ballistic scaling of mean-squared displacement in the $z$ direction demonstrating \textit{active} like dynamics, while the dynamics along the transverse ($x$) direction displays \textit{diffusive} behaviour with a strong dependence on the motion along $z$. To explain these unique characteristics, we developed a $2D$-\textit{Langevin} model of a confined elliptic particle experiencing an additional stochastic force along $z$ to account for the arbitrary jumps. The numerical results show excellent qualitative agreement with the experimental observations. Our findings should pave the way for the design of high-efficiency Brownian engines in air, besides stimulating new research in the emerging field of photophoretic trapping.

\end{abstract}

\section{Introduction}


Optical tweezers have revolutionized the study and understanding of Brownian motion -- both diffusive and active -- using colloidal probes trapped under controlled conditions \cite{gieseler2021optical,volpe2023roadmap}. While trapped Brownian probes in liquids display diffusive motion in typical experimentally resolvable time scales due to the large viscosity of liquids \cite{neuman2004optical}, active Brownian motion is easily discernible in motile bacteria, designed active Brownian particles such as Janus particles, vibrobots etc., which are subjected to forces generated from internal physio-chemical processes or external perturbations ~\cite{fodor2016far,bechinger2016active,ramaswamy2017active,vrugt2024review}. The time scales of these forces are different, with diffusive forces relaxing faster compared to the active forces, which can persist for significantly longer time. This usually leads to the near-ballistic scaling of mean squared displacement at a shorter lag time~\cite{howse2007self,lowen2020inertial}. On the other hand, particles trapped in air due to dipole forces display predominantly ballistic Brownian motion owing to the low viscosity of air \cite{li2010measurement,li2013brownian}. 
It is, however, unusual for the same particle optically trapped in a fluid to simultaneously display both types of Brownian motion, spontaneously arising without any external perturbations. Such motion, if experimentally realizable, could be very useful in designing Brownian engines with particularly high efficiencies, even exceeding the equilibrium Carnot limit~\cite{krishnamurthy2016micrometre,holubec2020active,nalupurackal2023towards,albay2023colloidal}.

Recently, photophoretic forces generated by loosely focused light beams incident on absorbing particles~\cite{shvedov2009optical,shvedov2010giant,sil2022trapping,pahi2023comparison} have introduced a new dimension in the trapping of absorbing microparticles in air. This force, originating from thermal effects, is several orders of magnitude stronger than the radiation pressure force, thereby eliminating the need for high numerical aperture lenses, as required in conventional optical tweezers. While photophoretic trapping has enabled diverse applications, such as 3D volumetric displays~\cite{smalley2018photophoretic} and engineered aerosols~\cite{keith2010photophoretic}, the dynamics of photophoretically trapped particles, especially the statistical nature of their fluctuations, remain largely underexplored. Thus, while confinement in the direction of the trapping beam ($z$) is understood to be a consequence of the balance of gravity and longitudinal photophoretic forces, an important question not yet fully addressed is the origin of trapping in the transverse ($x$) direction, where there appears to be a restoring force in action~\cite{bera2016simultaneous}.  This is even more relevant in the case of large particles (of diameter greater than 20 $\mu$m) that have a high inertial mass, and should therefore be very difficult to confine in all three dimensions. Furthermore, particle asymmetry appears to be an important factor in photophoretic trapping as well, both in terms of aiding the trapping~\cite{}, and in generating particle rotations~\cite{lin2014optical,chen2018temporal,pahi2024study}. An interesting area of exploration may therefore be the trapping of large asymmetric particles, and careful quantification of their Brownian dynamics in both $z$ and $x$ directions. 

In this paper, we focus on the trapping of single large asymmetrical microclusters of Carbon particles, and quantify their Brownian dynamics experimentally. We observe rather intriguing motion of the microclusters, manifested in the form of arbitrary jumps along the $z$ direction, possibly due to the interplay of gravitational and photophoretic forces. In addition, they also display fluctuations having significantly greater amplitude than typical thermal fluctuations along this direction. Interestingly, our earlier work with such clusters had revealed spinning and orbital motions \cite{pahi2024study} - however, in the present case,  we employ particles larger in size than those in Ref.~\cite{pahi2024study}, and do not observe any rotation at all. In contrast, the observed jumps in the $z$ direction occur at irregular intervals, resulting in a bimodal-shaped position distribution function and a near-ballistic scaling of the mean square displacement (MSD). These characteristics align with the dynamics of an active Brownian particle in harmonic confinement~\cite{caraglio2022analytic,nakul2023stationary,malakar2020steady,dauchot2019dynamics,buttinoni2022active}. Conversely, the dynamics along the $x$-direction remain diffusive over short timescales. Indeed, we believe that the simultaneous observation of such contrasting dynamics along two degrees of freedom within a single system is unprecedented, and can result in the controlled design of Brownian engines in air capable of very high efficiency (even higher than the Carnot limit) due to the presence of the active component in the dynamics~\cite{}. Additionally, we also demonstrate that our observed phenomena can be qualitatively understood using a simplified Langevin-based model with fit parameters gleaned from experiments. Our work can thus help motivate future research in various aspects of photophoretic traps in order to obtain a deeper understanding of the forces involved and the particle dynamics they can generate, paving the way for future applications and further advancements in this fascinating field.


The structure of the paper is as follows: Section 2 describes the details of the experiment, Section 3 presents the experimental results, Section 4 provides details of our phenomenological model, Section 5 presents the outcomes of the simulation, and Section 6 concludes the study.

\section{Experiment Materials and Methods}

\begin{figure}[H]
\centering
\includegraphics[width=\linewidth]{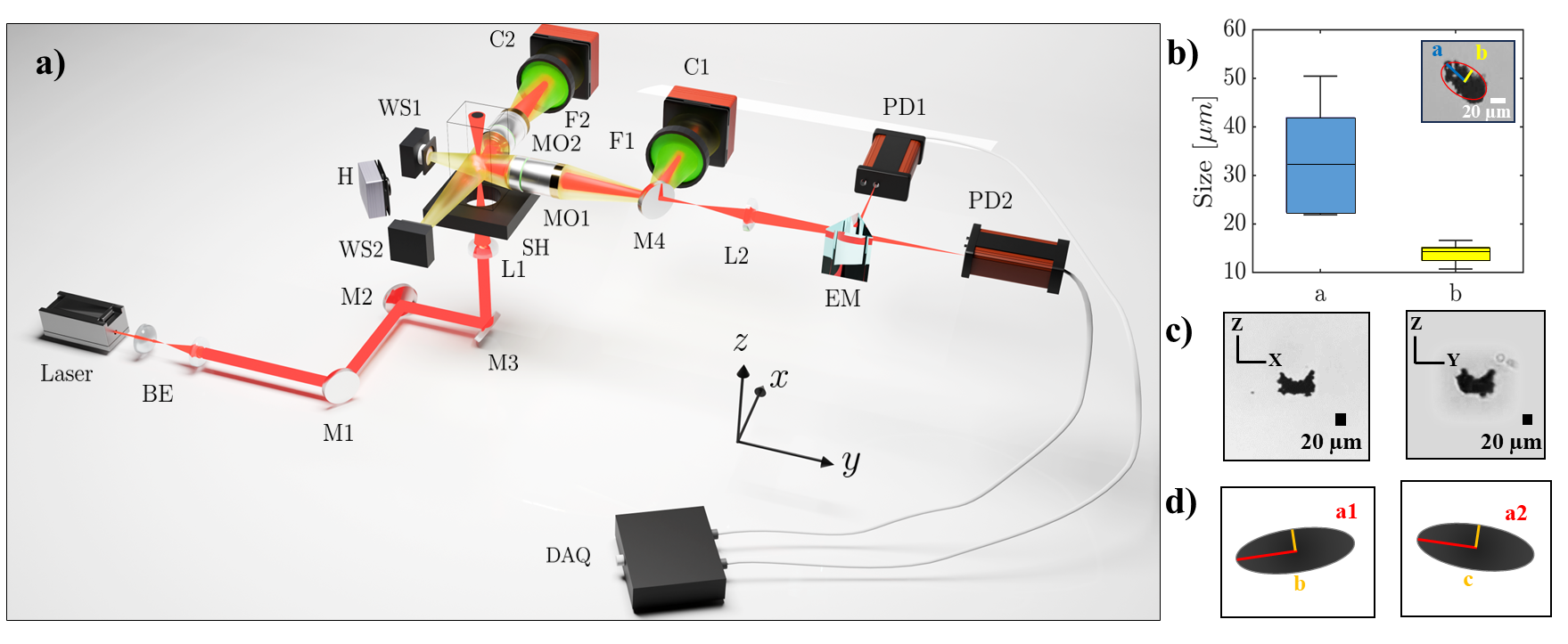}
\caption{(a) \textbf{Experiment schematic:} Labels denote the following, BE : Beam expander, M : Mirror, M4 is mounted on a flip mount,  L : Lens, SH : Sample holder, MO: Microscope objective, WS: White light source, F : Filter, C1 and C2 : Camera to capture $xz$ and $yz$ plane of the trapped microclusters respectively, DAQ : Data acquisition card, PD1 and PD2 : Photodiode 1 and photodiode 2, EM : Edge mirror (b) This shows the box and whisker plot of the semi major axis(a) and semi minor axis(b) of the trapped microclusters respectively. The inset at the top right side shows a representative trapped particle. (c) It shows the image of a trapped microcluster from two directions i.e. from $xz$ (left) and $yz$ (right) plane captured using C1 and C2, respectively. (d) This shows how we measure the major axis and minor axis of the trapped microcluster by considering the microcluster as an ellipsoid.}
\label{Fig1}
\end{figure}

In our experiment, we have used polydispersed carbon microspheres (sizes between 2-13 $\mu m$) from Sigma-Aldrich as trapped particles. 
These particles were trapped using a diode laser(from Oxius) of wavelength 640 nm. The laser beam is loosely focused using a 25 mm lens into the sample chamber with the focal plane at approximately the center of the sample chamber. The sample chamber is made by gluing together glass slides. Carbon microspheres are initially placed on the top glass lid of the sample chamber, and then loaded into the chamber with the help of a physical perturbation. The particle gets trapped at the position where the downward gravitational force balances the upward photophoretic force. The beam direction is vertically upwards (compared to the falling particles, as illustrated in Figure \ref{Fig1}), which we refer as $+z$ (upward) and $-z$ (downward). The $x$ and $y$ axes represent directions transverse to the beam axis. Now, a microcluster gets trapped at a power of 70 mW at the trapping region. However, at this power, the microcluster exhibits significant motion along the $z$-axis, leading to difficulties in detecting position fluctuations. The amplitude of $z$ motion decreases with laser power\cite{pahi2024study,sil2024ultrastable}, so that the laser power was reduced to levels where the scattered light remained within the detector's active area, typically around 20 mW. This power level was adjusted slightly for each microcluster to ensure optimal detection.

The scattered light from the $xz$ and $yz$ planes of the trapped microcluster is collected using two 10X Olympus microscope objectives. Each microscope objective is mounted on two linear translational stages: one for adjusting the position along the $z$-axis and the other for focusing the scattered light onto the image and detection planes. The scattered light corresponding to the $xz$ plane of the trapped particle is first directed to a mirror mounted on a flip mount, allowing the light to be sent either to the C1 camera or the balanced detector. C1 and C2 are CMOS cameras used to track the motion of the trapped microclusters along the $xz$ and $yz$ planes, respectively. Images from C1 and C2 are used to estimate mass of the trapped microclusters using the method described in this work\cite{sil2020study}, also elaborated in \textit{supplementary information}. The balanced detection setup comprises of two photodiodes placed perpendicular to each other, enabling the measurement of very small displacements by analyzing the difference in signals from the two photodiodes. We use the balanced detector to measure $x$ position fluctuation of the trapped microcluster\cite{pahi2024study,bera2017fast}. Hence, $x$ trajectory can be obtained from both the balanced detector and video analysis, but $z$ trajectory is obtained from video analysis. We calibrated the camera's field of view using a micrometer scale and determined a conversion factor from pixels to micrometers. Videos were collected at 26 frames per second. For balanced detection, data was acquired using a data acquisition card at a sampling frequency of 20 kHz for 100 seconds, with 50 dB gain. 




\section{Experimental Observations}

\begin{figure}[H]
\centering
\includegraphics[width=\linewidth]{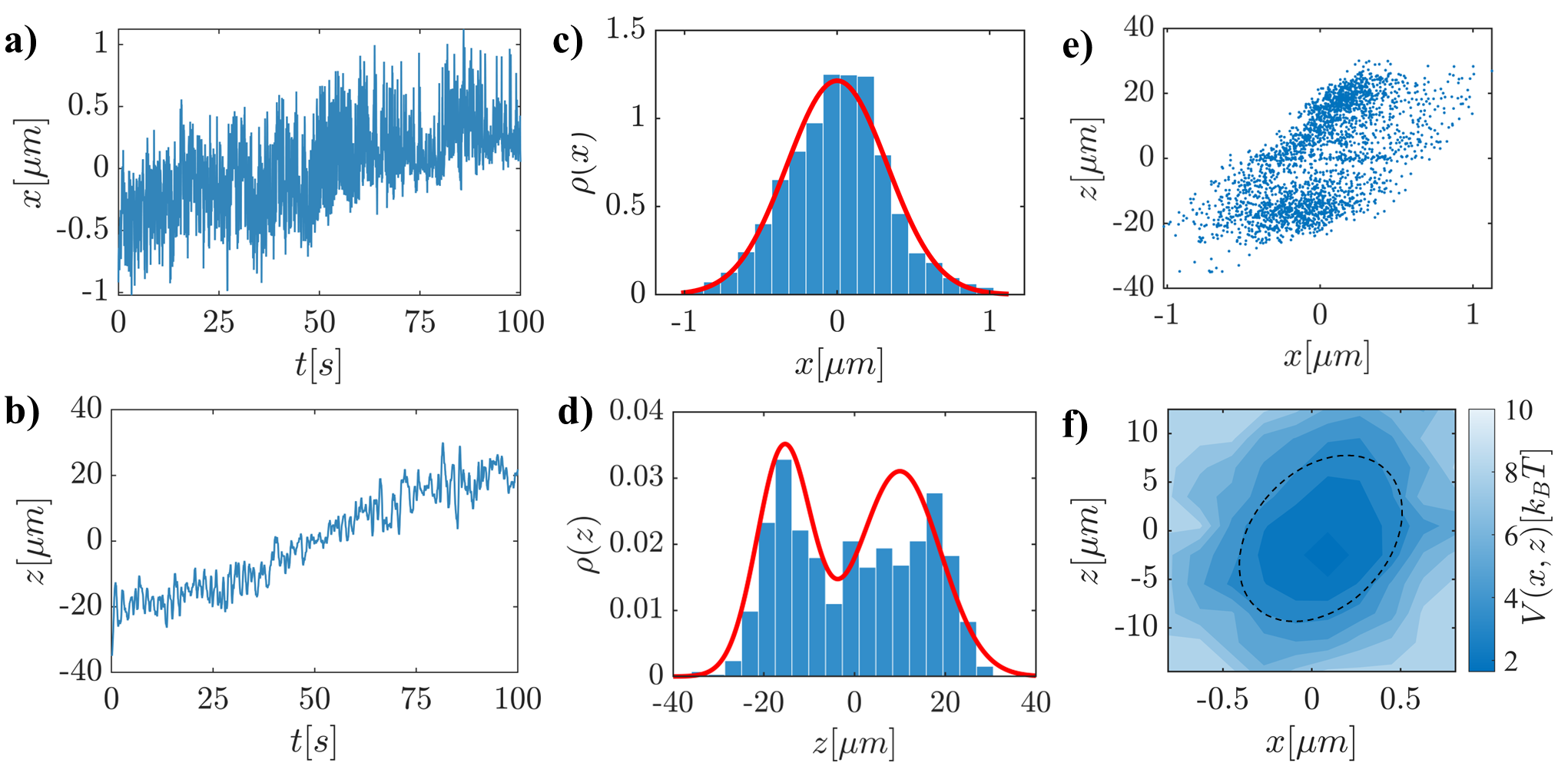}
\caption{\textbf{Experiment plots (1)} (a) Position fluctuations of a trapped microcluster along the $x$ direction.
(b) Position fluctuations of the same trapped microcluster along $z$. (c) The probability distribution function of $x$-position, exhibiting a Gaussian distribution. (d) The probability distribution function of the $z$ position, displaying a bimodal distribution. (e) The trajectory plot of the microcluster in the $xz$ plane (f) The potential as a function of $x$ and $z$ is tilted and is calculated from the detrended measured positions of the trapped microcluster.}
\label{Fig2}
\end{figure}

The mean subtracted time series along $x$ and $z$ are shown in Figure \ref{Fig2} (a-b). The trajectories clearly indicate that the dynamics along $x$ display a similar feature of upward drift as observed in the dynamics along $z$. This similarity of behaviour was consistently observed across all the datasets analyzed which arises due to the coupling between the two dynamic degrees of freedom ($x$ and $z$), owing to the asymmetric shape of the microcluster.  We find that the probability distribution function along $x$ is typically Gaussian (Figure \ref{Fig2} (c)) with a small skew due to the drift, whereas the probability distribution function along $z$ displays a bimodal nature, as shown in Figure \ref{Fig2} (d). From the measured position fluctuations along $x$ and $z$, we aim to determine the potential landscape in the $xz$ plane. To accurately extract the potential, we first detrend the trajectory to remove any systematic variations arising from particle dynamics. Further details on the potential calculation can be found in the \textit{Supplementary Information}.
Note that the underlying potential can be well-approximated as a tilted $2D$- harmonic potential as shown in Figure \ref{Fig2}(f), which indicates that the coupling is present in the potential itself.

\begin{figure}[H]
\centering
\includegraphics[width=\linewidth]{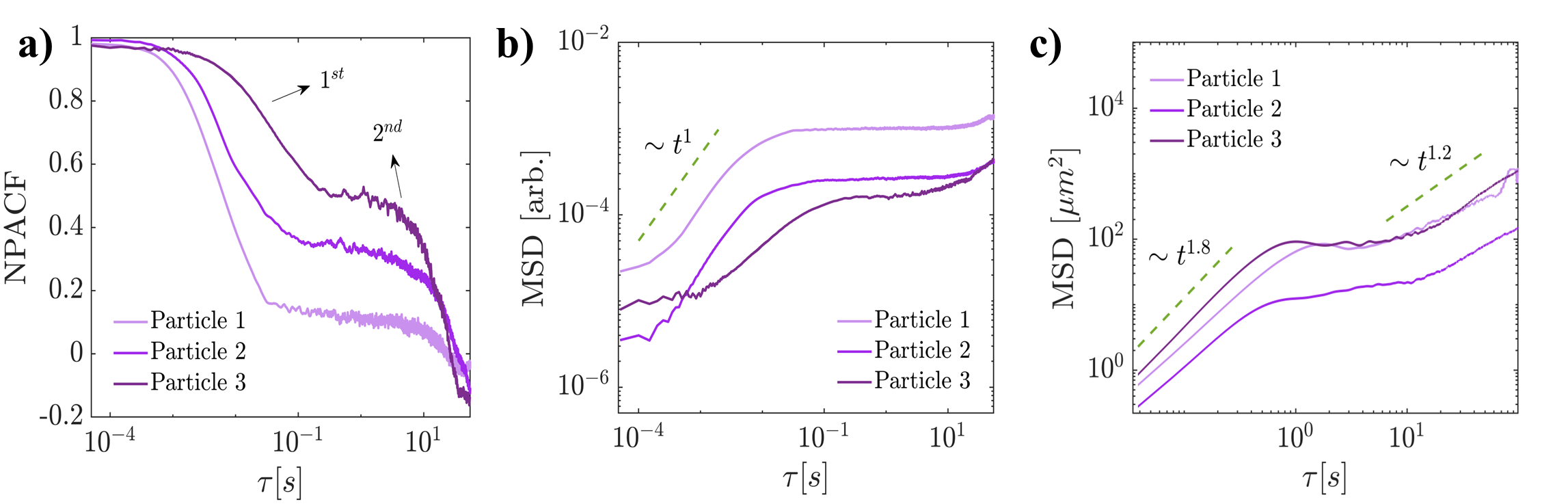}
\caption{\textbf{Experiment plots (2)} (a) Normalised position autocorrelation function of three trapped particles along $x$ direction exhibiting a two step decay. (b) Mean squared displacement of three trapped particles along $x$ showing diffusive scaling (c) Mean squared displacement along $z$ having two scalings as shown. The three particles differ in mass, with particle 1, particle 2, and particle 3 having masses of 19 , 30, and 67 pico-kgs, respectively.}
\label{Fig3}
\end{figure}

We then compute the mean squared displacement(MSD) and normalised position autocorrelation function (NPACF) from the measured position fluctuations. We define, 
MSD ($\tau$) = $\langle (x(t+\tau) - x(t))^2\rangle$ and NPACF ($\tau$)= $\frac{\langle (x(t+\tau) - \bar{x})(x(t)-\bar{x})\rangle}{\text{var}(x)}$ where $\tau$ represents the lag time and $\bar{x}$ is the mean of $x$. We used the $x$ position data collected using the balanced detector (shown in \textit{supplementary information}) at a rate of 20 $kHz$ to calculate the NPACF and MSD along $x$ as the corresponding time scales are much smaller compared to the motion along $z$, which warrants much faster data collection. The position autocorrelation function along $x$ shows two-step decays, as shown in Figure \ref{Fig3} (a). The scaling for mean squared displacement along $x$ at shorter lag is $\sim t^{1}$ (Figure \ref{Fig3} (b)) i.e. diffusive, which is surprising for a particle trapped in air, where the Brownian motion is typically ballistic due to the low viscosity~\cite{}. This actually suggests that the particle experiences a higher drag force along the $x$-axis originating from the effective viscosity being higher than that of air. Indeed, this may be the reason behind the stable radial confinement of particles in photophoretic traps using loosely focused Gaussian beams generated by lenses having much lower NA than that required for optical traps created by dipole forces. This confinement is demonstrated by a plateau in the MSD. 

As observed from the $z$ trajectory in Figure \ref{Fig2} (b), as well as Video 1 given in {\it Supplementary Information}, it is clear that the particle experiences arbitrary jumps at irregular intervals along the beam direction. We attribute these kicks to the interplay of gravitational and photophoretic forces along $z$. As a result, there is a drift in the $z$ trajectory. The MSD obtained along $z$ shows ballistic-like scaling of $\sim t^{1.8}$ at shorter lags and near diffusive-type scaling of $\sim t^{1.2}$ at later times (Figure \ref{Fig3} (c)). Note that the bimodal nature of position probability distribution and near-ballistic behaviour in the MSD are characteristic features of active particles in a harmonic trap \cite{shen2019far, caraglio2022analytic}. Interestingly, the activity of these particles can be controlled by varying the laser power (Figure \ref{Fig6}(a)), which we elaborate later on. 

We develop a phenomenological model, which includes key experimental details to understand these intricate experimental observations and the underlying dynamics. The model and its analysis are explained in detail in the following sections.


\section{Phenomenological model}

\begin{figure}[H]
\centering
\includegraphics[width=\linewidth]{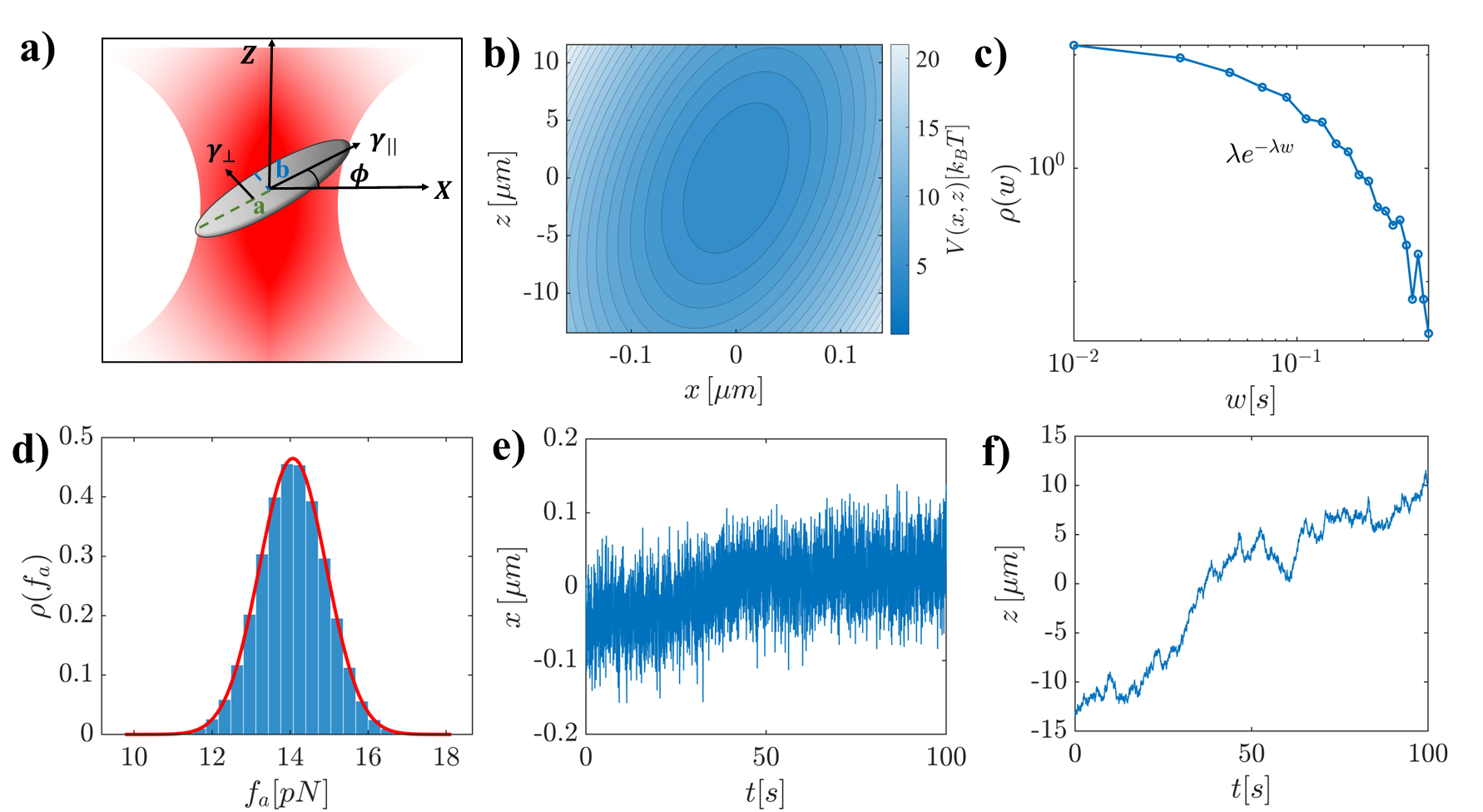}
\caption{\textbf{Simulation plots (1)} (a) This is a representation of a trapped ellipsoidal particle oriented at an angle $\phi$ relative to the $x$-axis, with $a$ and $ b$ denoting the semi-major and semi-minor axes, respectively. The parallel and perpendicular drag coefficients,$\gamma_\parallel$  and $\gamma_\perp$, are also illustrated. (b) Coupled harmonic potential along $x,z$ (c) Normalised exponential probability distribution of waiting time. Here, $w$ represents the waiting time. (d) The strength of the additional force ($f_{\text{add}}(t)$) having a Gaussian distribution with non-zero mean (e) $x$-time series (f) $z$-time series.}
\label{Fig4}
\end{figure}

We provide a 2D Langevin-based phenomenological model to explain the dynamics of the trapped particles in the $xz$ plane. For simplicity, we consider these asymmetric particles as ellipsoids with an aspect ratio similar to the trapped particle (Figure \ref{Fig4} (a)). 

The confining potential in the $xz$ plane is modelled as a coupled harmonic potential of the form, 
\begin{equation}
    V(x, z) = \frac{1}{2} (k_x x^2 + k_z z^2) - \alpha xz
    \label{potential}
\end{equation}
with $k_{x}$ , $k_{z}$ being the stiffness constant along $x$, $z$ respectively and $\alpha$ is considered to be a coupling constant. The potential structure is the same as that obtained from the experiment as represented in Figure \ref{Fig4} (b). The equilibrium dynamics of a microscopic ellipsoidal particle of mass $m$ in a confining potential $V$ can be represented as:
\begin{equation}
    m \dot{v}_i(t) = - \sum_{j} \gamma_{ij}(\phi) v_j(t) - \frac{\partial V(i, j)}{\partial i} + \xi_{v_i}(t), \quad i,j = (x,z) 
    \label{langevin}
\end{equation}
with the thermal forces following certain statistical features: $\langle \xi_{v_i}(t) \rangle = 0$ and $\langle \xi_{v_i}(t) \xi_{v_j}(t') \rangle = 2k_B T \gamma_{ij} \delta(t - t')$, where $k_B$ is the Boltzmann constant and $T$ is the temperature of the bath.

The drag coefficient $\gamma_{ij}$ of the particle in the air is anisotropic due to the elliptic shape of the particle. It can be defined as~\cite{han2006brownian,dutta2023microscopic}:
\begin{align}
    \gamma_{ij} &= 
   \begin{pmatrix}
       \gamma_{xx} & \gamma_{xz}\\
       \gamma_{zx} & \gamma_{zz} 
   \end{pmatrix}
     = 
   \begin{pmatrix}
   \gamma_\parallel \cos^2 \phi + \gamma_\perp \sin^2 \phi & (\gamma_\parallel - \gamma_\perp) \sin \phi \cos \phi \\
   (\gamma_\parallel - \gamma_\perp) \sin \phi \cos \phi & \gamma_\parallel \sin^2 \phi + \gamma_\perp \cos^2 \phi
   \end{pmatrix}
\end{align}
where, $\phi$ is the orientation angle of the particle, measured as the angle between the long axis of the particle and the horizontal axis of the lab frame (Figure\ref{Fig4}(a)), and $\gamma_\perp = 6 \pi \eta_\perp a$, $\gamma_\parallel = 6 \pi \eta_\parallel b$, with $a, b$ being the lengths of the semi-major and semi-minor axes of the projection of the ellipsoidal particle in the $xz$ plane. $\eta_\parallel$ and $\eta_\perp$ are the viscosity experienced by the particle along a direction parallel and perpendicular to the long axis of the particle, respectively. 

Now for our system, the asymmetric-shaped microcluster is observed to fluctuate with a larger amplitude in the $z-$ direction compared to that in the $x-$ direction. The dynamical equations corresponding to such motion of the trapped particle in the $xz$ plane are written as:


\begin{equation}
    \begin{aligned}
        m \ddot{x}(t) &= -(\gamma_{xx} \dot{x}(t) + \gamma_{xz} \dot{z}(t)) - k_x x(t) + \alpha z(t) + \xi_{v_x}(t)\\
        m \ddot{z}(t) &= -(\gamma_{zx} \dot{x}(t) + \gamma_{zz} \dot{z}(t)) - k_z z(t) + \alpha x(t) + \xi_{v_z}(t) + f_{\text{add}}(t),
    \end{aligned}
\end{equation}
with an additional force $f_{\text{add}}(t)$ along $z$- direction. 
The additional force is considered as abrupt kicks to account for the irregular jumps observed from the experiment and it is defined as,
\begin{equation}
    f_{\text{add}}(t) = f_a \delta(t - t_n),
\end{equation}
where a kick occurs at time $t_n$ and the time between successive kicks, called the waiting time $w$ (where $w(n) = t_n - t_{n-1}$) follows a normalised exponential distribution $\lambda e^{-\lambda w}$ where $\lambda$ being the rate of kicks. This distribution is shown in Figure \ref{Fig4} (c). 
The strengths of such kicks ($f_a$) are sampled from a Gaussian distribution with a non-zero mean (Figure \ref{Fig4} (d)) to account for the drift in $z$ time series and variance proportional to additional temperature $\delta T$, such that 
\begin{equation}
\begin{aligned}
    \langle f_a(t) \rangle & = c \\
    \langle f_a(t) f_a(t') \rangle & = 2 k_B \delta T \gamma_{zz} \delta(t - t')
\end{aligned}
\end{equation}
Here, $\delta T$ accounts for the additional thermal fluctuations arising from the interplay of the photophoretic force and gravity. Since the additional force is additive along the $z$ direction, the particle effectively experiences thermal fluctuations corresponding to a higher temperature $(T + \delta T)$ at irregular intervals.

Note that a similar type of abrupt forcing mechanism has recently been used to describe the dynamics of a harmonically confined particle within a bath of active particles \cite{di2024brownian}, where random kicks arise due to interactions with the active particles. These stochastic kicks are characterized as Poisson shot noise (PSN) and the number of kicks in a finite interval follows a Poissonian distribution.

\section{Numerical Results and Discussion}

The dynamical equations describing the model are numerically solved by discretizing with a time interval $\Delta t = 50\mu s$ (which is less than all the other timescales of the system), following the 1st-order Euler-Maruyama algorithm. The details of the numerical simulation are provided in the \textit{Supplementary Information}. To obtain the diffusive nature of the mean squared displacement (MSD) along \( x \), which suggests that the particle experiences a higher effective viscosity along \( x \) than that of air, the parallel viscosity \( \eta_{\parallel} \) was taken to be 56 times greater than the viscosity of air, in order to match experimental results. We explicitly show the scaling dependence of MSD along $x$ estimated from numerical trajectories by varying the parallel viscosity, \( \eta_{\parallel} \), in the \textit{Supplementary information}. Apart from this, most of the parameters used in the model were obtained from the experiment (Table 1 of the \textit{supplementary information}).


\begin{figure}[H]
\centering
\includegraphics[width=\linewidth]{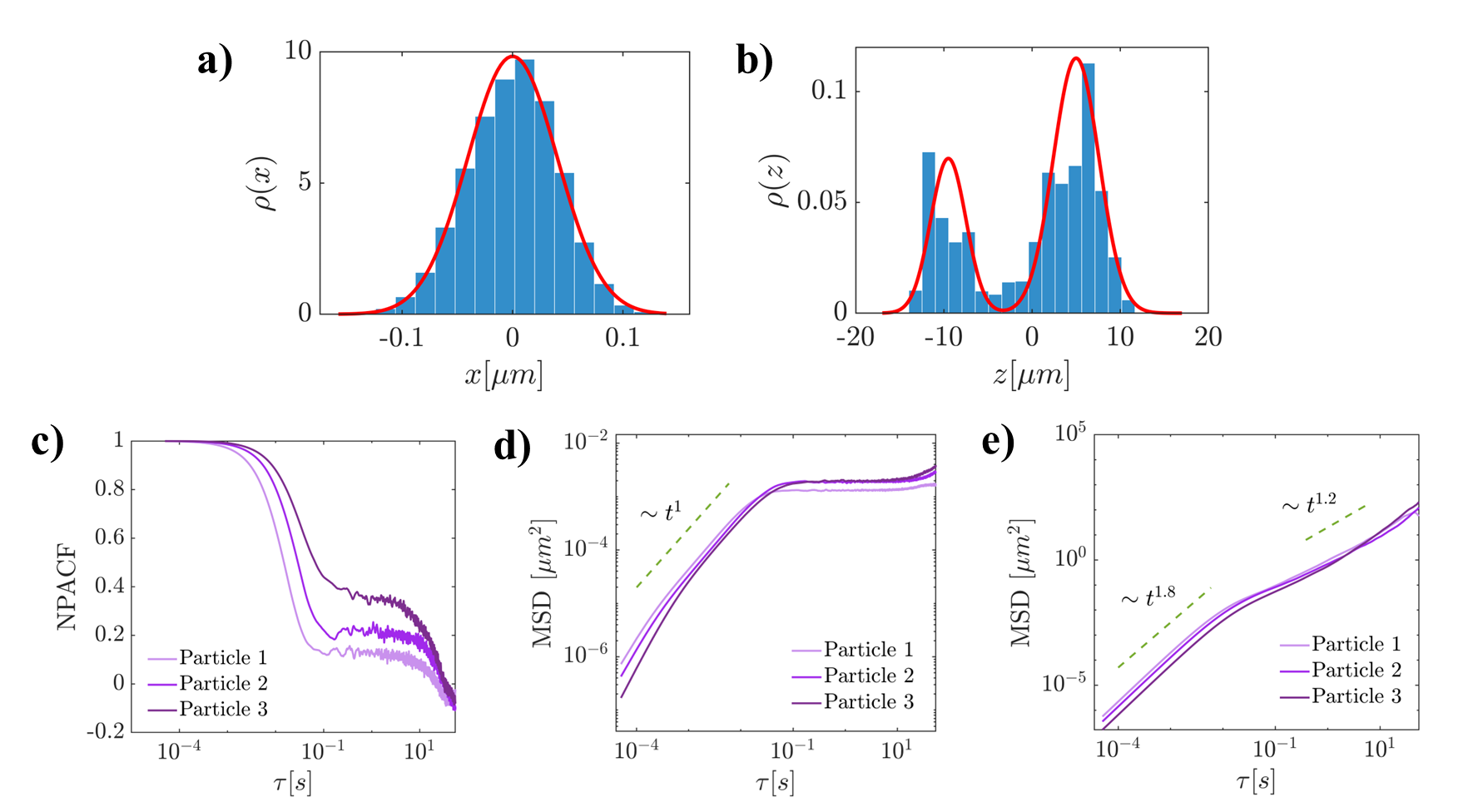}
\caption{\textbf{Simulation plots (2)} (a) Position probability distribution function along $x$. (b) Position probability distribution function along $z$.(c) Normalised position autocorrelation function along $x$ for three different particles. (d) Mean squared displacement along $x$. (e) Mean squared displacement along $z$. The three particles have masses of 19, 30, and 67 pico-kgs, with corresponding stiffness values along $x$ of $6.14\ pN/\mu m$, $4.26\ pN/\mu m$ and $4.15\ pN/\mu m$, respectively.}
\label{Fig5}
\end{figure}
The nature of numerical trajectories along $x$ and $z$ direction match reasonably well with our experimental observations (see Figure \ref{Fig4}(e)-(f)), where the trend in trajectory along $x$ and $z$ are similar. The corresponding probability distribution function along $x$- direction remains Gaussian (Figure \ref{Fig5}(a)), while it clearly appears bimodal along $z$-direction (Figure \ref{Fig5}(b)). Additionally, the normalised position autocorrelation function (NPACF) of the trajectories along $x$- direction also bears the signature of multiple decays following the dynamics along $z$ that we observe in the experiment as well (Figure \ref{Fig5}(c)). Note that NPACF along $x$ with such multiple decay structure appears only in the presence of arbitrary kicks along the $z$- direction. Importantly, the features of the MSDs along both directions display similar characteristics (at a short time, MSDx:$\sim t^1$ and MSDz:$\sim t^{1.8}$ ) computed from the experimental trajectories (Figure \ref{Fig5}(d)-(e)).

\begin{figure}[H]
\centering
\includegraphics[width=\linewidth]{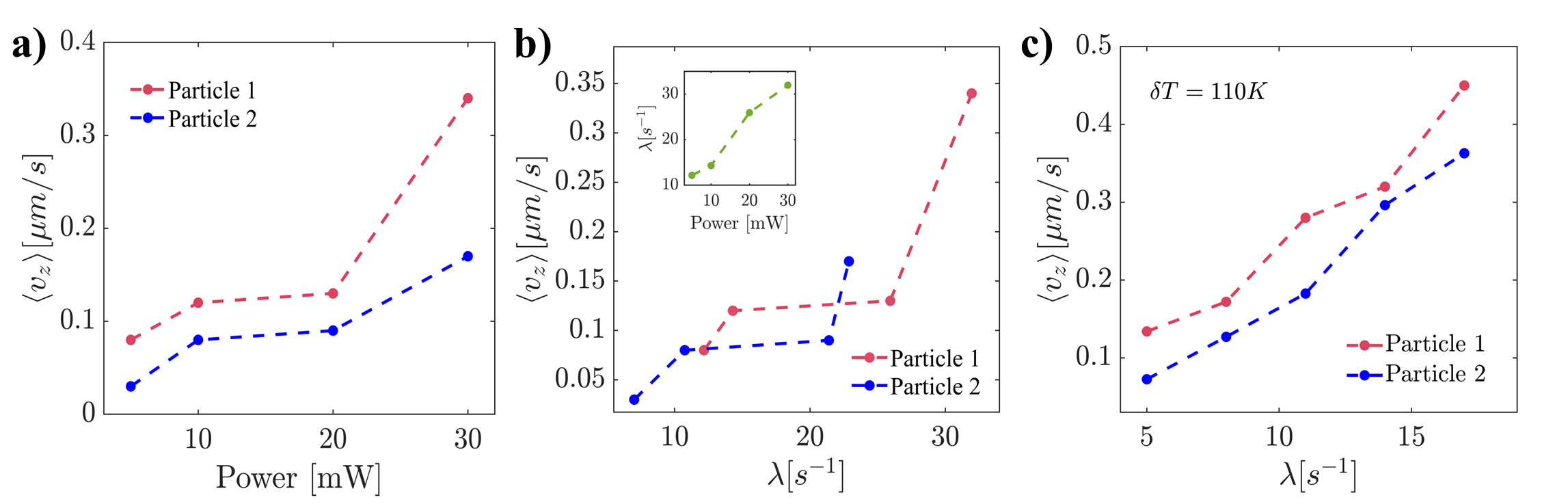}
\caption{ (a) Variation of mean $z$ velocity with power at the trapping region (b) Variation of $z$ velocity with rate of kicks ($\lambda$) for two different trapped particles. The inset shows the variation of $\lambda$ with laser power at the trapping region for one of the trapped particles. (c) Dependence of mean $z$- velocity ($\langle v_{z} \rangle$) with rate of kicks($\lambda$) estimated from numerical trajectories.}
\label{Fig6}
\end{figure}

\textit{Experimental control over activity.-}
Next, we focus on the velocity of the particle along $z$ to understand the inherent active feature embedded in the dynamics.  The velocity time series along the $z$-axis is computed as $v_z(t) = (z_t - z_{t-1})/dt$ - from the $z$-position data. We first observe that the mean $z$- velocity of the particle, $\langle v_z \rangle$ is enhanced with increasing laser power (Figure~\ref{Fig6}(a)), indicating the possibility of controlling the activity of the particle by changing the laser power. To understand the origin of the active features along $z$-direction, we further determine the rate of kicks ($\lambda$) from the experimental trajectories. To do that, the calculated velocity fluctuations are compared with the equilibrium standard deviation of the velocity fluctuations i.e. $\sigma_{v_z}^{eq} \sim \sqrt{\frac{k_{B}T}{m}}$. 
A kick is identified when the velocity amplitude exceeds $\sigma_{v_z}^{eq}$. The corresponding time instants ($t_n$) are recorded, and the waiting time between successive kicks is estimated as $w(n) = t_n - t_{n-1}$.
The distribution of waiting times extracted from the experimental trajectories is then fitted to the same exponential function, $\lambda e^{-\lambda w}$ - used in the model, to extract the parameter $\lambda$.  We also verify this method of extracting the rate of kicks ($\lambda$) with numerical trajectories for which $\lambda$ is known \textit{a priori}.


Interestingly, the rate of kicks ($\lambda$) obtained from the experimental trajectories is also found to increase at higher laser powers, resulting in higher mean velocity (Figure~\ref{Fig6}(b)). A similar dependence between $\langle v_z \rangle$ and $\lambda$ obtained from numerical trajectories as shown in Figure~\ref{Fig6}(c) further validates our observations.

It is now obvious that the system in the 2D phase space ($xz$ plane) stays inherently away from equilibrium, and the irregular kicks along $z$ act as the source of non-equilibrium behaviour. Specifically, the spontaneous active-like dynamics along one direction ($z$) drive the dynamics of the other degree of freedom ($x$). Most importantly, the `activity' levels can be externally tuned as well. This leads us to intuitively speculate that this model should be able to provide a unique opportunity to design Brownian work-to-work converter machines with different operational modes, with their thermodynamic properties~\cite{gupta2017stochastic,manikandan2019efficiency,das2022inferring} open to exploration within the framework of stochastic thermodynamics~\cite{seifert2012stochastic,seifert2019stochastic}.  

\section{Conclusion}

In this study, we experimentally investigate the fluctuation dynamics of single large asymmetric microcluster trapped in air. Our findings reveal that the measured dynamics display simultaneous active and diffusive behaviour, with the active behaviour displayed in the direction of particle motion along the trapping laser beam caused by the interplay between photophoretic forces and gravity. This leads to irregular jumps of the microcluster in that direction, causing a near-ballistic time scaling of the MSD of the positional variance and bimodal probability distributions. To model these dynamics, we develop a phenomenological model based on the two-dimensional Langevin equation, making to the best of our knowledge, the first attempt of this kind in this research area. Thus, our model captures the experimental observations with qualitative accuracy, and provides deeper insight into the fundamental forces governing the system. Indeed, the whole system is fundamentally novel and intriguing as the two motional degrees of freedom possess mutually contrasting dynamic behaviours. Moreover, the activity of the process along $z$-direction can also be controlled through the power of the loosely focused Gaussian beam that is used to confine the particle, providing an experimentalist a simple handle to tune the activity. Note that the system inherently stays out of equilibrium due to abrupt jumps along the beam propagation direction, promising very interesting applications in the design of Brownian engines in air with efficiency even higher than the Carnot limit due to the activity and non-equilibrium-ness of the system~\cite{krishnamurthy2016micrometre,albay2023colloidal}. In this context, the thermodynamic aspects of this system in terms of trajectory energetics (work, heat, entropy) could also show enigmatic features ~\cite{manikandan2021quantitative,das2022inferring,das2023enhanced,das2024irreversibility,lyu2024learning} -- which would require further exploration using the tools of \textit{stochastic thermodynamics}~\cite{seifert2012stochastic,seifert2019stochastic}. Additionally, since the dynamics observed in such systems are often too complex to solve analytically, it offers an opportunity to develop or utilise parameter estimation techniques~\cite{bera2017fast,singh2018fast,gerardos2025principled} to understand the intricacies of the stochastic process. Future work can also explore a broader range of absorbing materials, particle shapes, and aspect ratios to further characterize and generalize the observed active-like behaviour. We intend to report interesting research results in these directions in the near future.

\begin{acknowledgement}

The authors acknowledge Indian Institute of Science Education and Research (IISER) Kolkata for the funding and facilities.
BD is thankful to the Ministry Of Education of Government of India for financial support through the Prime Minister’s Research Fellowship (PMRF) grant.
\end{acknowledgement}

\begin{suppinfo}

The Supporting Information is available free of charge in the journal website. The supplementary information contains more details on the experimental and numerical analysis along with videos showing motion of trapped particles.

\end{suppinfo}

\providecommand{\noopsort}[1]{}\providecommand{\singleletter}[1]{#1}%
\providecommand{\latin}[1]{#1}
\makeatletter
\providecommand{\doi}
  {\begingroup\let\do\@makeother\dospecials
  \catcode`\{=1 \catcode`\}=2 \doi@aux}
\providecommand{\doi@aux}[1]{\endgroup\texttt{#1}}
\makeatother
\providecommand*\mcitethebibliography{\thebibliography}
\csname @ifundefined\endcsname{endmcitethebibliography}  {\let\endmcitethebibliography\endthebibliography}{}

\section*{Supplementary information : Simultaneous active and diffusive behaviour of asymmetric microclusters in a photophoretic trap}


\section{Measurement of the mass of the trapped particles}

We use the particle's image obtained from C1 and C2, which capture the projection of the trapped particle along the $xz$ and $yz$ planes, respectively, to estimate its mass. We approximate the trapped particle as an ellipsoid, so its volume is given by  $V = \frac{4\pi abc}{3}$. The semi-major axes ($a_1$ and $a_2$) are determined using ImageJ software from the particle’s projections in the $xz$ and $yz$ planes respectively. The projected area along both planes is measured using ImageJ, and the semi-minor axes $b$ and $c$ are obtained by applying the formula,  $A1 = \pi a_1 b \quad\text{and}\quad A2 = \pi a_2 c$   respectively (For notation refer to Figure 1(d) of the main text). Here $A1$ and $A2$ are the areas measured in $xz$ and $yz$ plane. The value of $a$ is then determined by averaging $a_1$ and $a_2$. Finally, the mass of the particle is estimated by multiplying the calculated volume with the density of carbon microspheres ($1.95 \, \text{g/cm}^3$), as specified by Sigma Aldrich (for more details of this method, refer to supplementary Information of Ref. \cite{sil2020study}). 


\section{Evaluating the potential at the trapping region}

The form of the potential along $xz$ is evaluated from the trajectory of the trapped particle extracted from video analyses of its motion. The centroid of the particle is tracked frame by frame to extract the $x$ and $z$ positions of the particle over time. Both $x$ and $z$ trajectories have particular trends (Figure 2. a-b) due to the motion of the particle caused by the interplay of photophoretic and gravitational forces. To remove the effect of these extra forces on the potential, the $x$- and $z$-position data is detrended to a degree such that there exists no visible trend in the trajectory. A two-dimensional normalized probability distribution function (PDF) with $n$ bins is computed for  the $x$- and $z$-positions of the trapped particle. The binned counts ($N_p$) are obtained using the \texttt{histcounts2} function in Matlab, resulting in a matrix of size $n\times n$, where each entry represents the number of counts in a specific bin. The midpoints of the bins for $x$ and $z$ are also calculated.

The potential energy  was derived from the normalized PDF using the Boltzmann relation:
\begin{equation}
E_{\text{M}} = -k_{B}T\ln(N_p),
\end{equation}
where $k_{B}$ is the Boltzmann constant, and T is the surrounding temperature.
The resultant potential energy landscape is visualized using a contour plot, with the $x$-bin midpoints, $z$-bin midpoints, and $E_{M}$( Figure 2(f) of main manuscript).


\section{X displacement time series obtained from balanced detection}

\begin{figure}[H]
    \centering
    \includegraphics[width=0.9\linewidth]{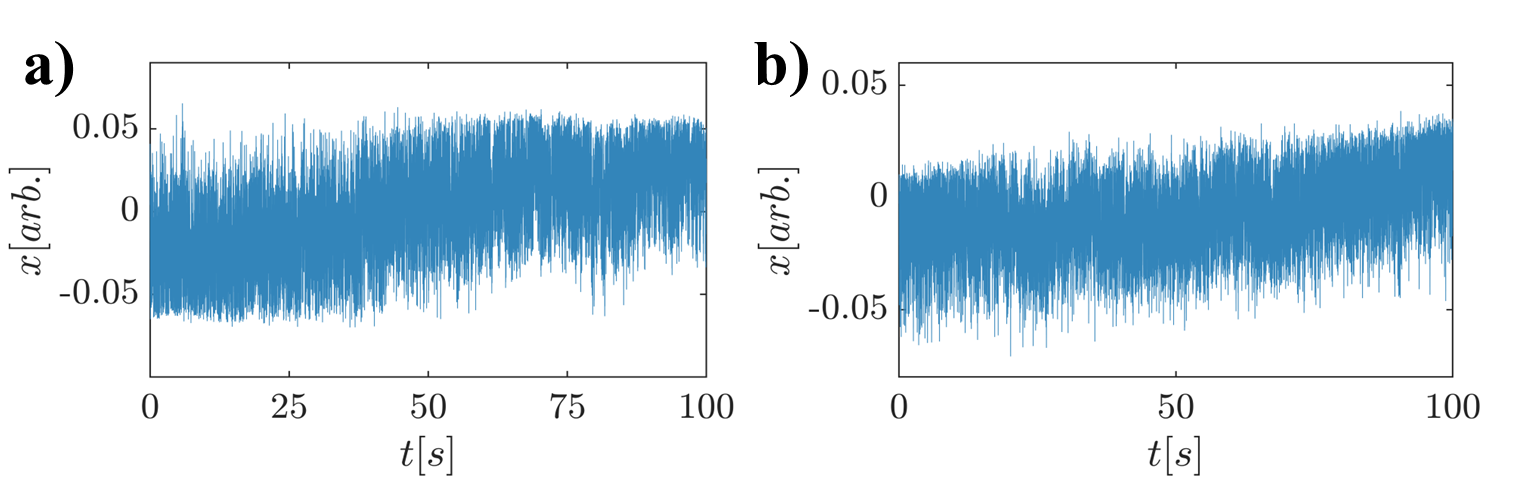}
    \caption{(a) and (b) shows $x$ time series of two trapped particles.}
    \label{x_pd}
\end{figure}

The $x$ position fluctuations are measured using a balanced detector at a sampling frequency of 20 kHz over a 100s time interval and are used to compute NPACF (Position autocorrelation function) and MSD (Mean squared displacement). Two such time series are shown in Figure\ref{x_pd}(a) and Figure\ref{x_pd}(b).

\section{Simulation details}

Equation 4 from the main manuscript can be written in matrix form as given below, with \(\dot{x}(t) = v_x(t); \, \dot{z}(t) = v_z(t)\), where $x$, $z$ are positions along the transverse and longitudinal (with respect to the laser beam) direction, while and $v_{x}$, $v_{z}$ are the velocities of the particle along $x$ and $z$, respectively:

\[
\begin{pmatrix}
\dot{x}(t) \\ 
\dot{z}(t) \\ 
\dot{v}_x(t) \\ 
\dot{v}_z(t)
\end{pmatrix}
=
-\begin{pmatrix}
0 & 0 & -1 & 0 \\
0 & 0 & 0 & -1 \\
\frac{k_x}{m} & \frac{\alpha}{m} & \frac{\gamma_{xx}}{m} & \frac{\gamma_{xz}}{m} \\
\frac{\alpha}{m} & \frac{k_z}{m} & \frac{\gamma_{zx}}{m} & \frac{\gamma_{zz}}{m}
\end{pmatrix}
\begin{pmatrix}
x(t) \\ 
z(t) \\ 
v_x(t) \\ 
v_z(t)
\end{pmatrix}
+
\begin{pmatrix}
0 \\ 
0 \\ 
\xi_{v_x}(t) \\ 
\xi_{v_z}(t)
\end{pmatrix}
+
\begin{pmatrix}
0 \\ 
0 \\ 
0 \\ 
f_{add}(t)
\end{pmatrix}
\]
The above matrix can be written in a single equation form as below:
\[
\Rightarrow \dot{\mathbf{X}}(t) = -\mathbf{F} \cdot \mathbf{X}(t) + \boldsymbol{\xi}(t) + \mathbf{F}_{\text{ext}}(t)
\]
where $\mathbf{X}$ matrix represents the following:
\[
\mathbf{X}(t) = [x(t), z(t), v_x(t), v_z(t)]^\top
\]
$\boldsymbol{\xi}(t)$ represents the noise matrix.
\[
\langle \boldsymbol{\xi}(t) : \boldsymbol{\xi}(t') \rangle = 2 \mathbf{D} \delta(t - t')
\]
The diffusion matrix can be defined as:
\[
D_{ij} = \frac{k_B T \gamma_{ij}}{m^2}.
\]
where $i, j = (x, z)$
\[
\mathbf{D} = 
\begin{pmatrix}
0 & 0 & 0 & 0 \\
0 & 0 &
0 & 0 \\
0 & 0 & D_{xx} & D_{xz} \\
0 & 0 & D_{xz} & D_{zz}
\end{pmatrix}
\]
Using Cholesky decomposition, the diffusion matrix can be decomposed as follows:
\[
\mathbf{D} = \frac{1}{2} \mathbf{G} \mathbf{G}^\top
\]
where (\(\mathbf{G}\)) is the noise strength matrix:
\[
\mathbf{G} =
\begin{pmatrix}
0 & 0 & 0 & 0 \\
0 & 0 & 0 & 0 \\
0 & 0 & \sqrt{2D_{xx}} & 0 \\
0 & 0 & \frac{\sqrt{2}}{\sqrt{D_{xx}}} D_{xz} & \sqrt{\frac{2(D_{xx}D_{zz} - D_{xz}^2)}{D_{xx}}}
\end{pmatrix}
\]
Finally, the experimentally obtained parameter values along with the range of values used in the simulation are tabulated below.

\begin{table}[H]
\centering
\begin{tabular}{|l|c|c|}
\hline
\textbf{Parameter} & \textbf{Simulation} & \textbf{Experiment} \\ \hline
$m$ (mass) & $(30-100)$ pico kg & same \\ \hline
$k_{x}$ & $(3-7) \times 10^{-6}$ N/m& same \\ \hline
$k_{z}$ & $(1-3) \times 10^{-10}$ N/m& same \\ \hline
$\alpha$ & $(1-3)\times 10^{-8}$ N/m& Not measurable \\ \hline
$\eta_\parallel (\text{viscosity parallel to semi major axis)}$ & $(50-70) \times \eta_{air}$ N.s/m$^2$  & Not measurable \\ \hline
$\eta_\perp(\text{viscosity perpendicular to semi major axis})$ & $1.57 \times 10^{-5}$ N.s/m$^2$ & Not measurable \\ \hline
$b (\text{semi minor axis})$ & $(12-15) \: \mu$m & same \\ \hline
$a(\text{semi major axis})$ & $(20-45) \: \mu $ m & same \\ \hline
$\phi$  & $(4-10)^\circ$  & same \\ \hline
$\delta T$ & $(80-120) $K & Not measurable \\ \hline
$\lambda$ & $(8-17)$ $s^{-1}$ & $(5-30)$ $s^{-1}$ \\ \hline
$\mu_{\text{kick}}$ & $1.5 \times 10^{-1}$ m/s$^2$ & Not measurable \\ \hline
\end{tabular}
\caption{Comparison of simulation and experiment parameters}
\label{model parameters}
\end{table} 
$k_{x}$ and $k_{z}$ are obtained by fitting the power spectral density with lorentzians with and without the mass term in the Langevin equation, respectively. This choice is based on the MSD scaling behavior of the trapped microclusters, which exhibit diffusive motion along $x$ and near-ballistic motion along $z$.

Note that in the presence of arbitrary kicks, the dynamics of asymmetric particles are coupled in all degrees of freedom,  which makes it difficult to extract all the parameters from experimental trajectories.  However, we believe that parameter-inference techniques based on simulations can be employed in future to understand the dynamics of such a system quantitatively -- it appearing rather too complex to solve analytically. Clearly, both approaches are well beyond the scope of our present work. 

\section{MSD of $x$-trajectories with varying viscosity}
In the main text, we mentioned that the dynamics of the asymmetric-shaped particle along $x$- direction are found to be diffusive as the MSD estimated from the experimental trajectories possess $\sim t^{1}$ scaling at short times (Figure 3(b) of main text).  Here, we show that we can get a similar feature from the numerical trajectories by considering the viscosity along $x$- direction, i.e. $\eta_\parallel$ to be higher than that of the air. 
\begin{figure}[H]
    \centering
    \includegraphics[width=0.5\linewidth]{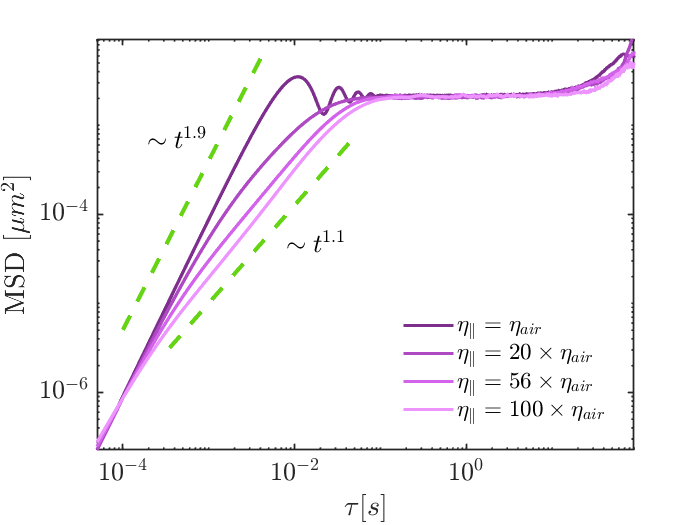}
    \caption{MSD along $x$- direction estimated from numerical trajectories with varying viscosity}
    \label{msdx_eta_varied_4_paper_plot}
\end{figure}


\end{document}